\documentclass[preprint,showpacs,showkeys,preprintnumbers,amsmath,amssymb]{revtex4}
\usepackage{amssymb}
\usepackage[dvips]{graphicx}

\begin{document}

\title{New interaction solutions of Kadomtsev-Petviashvili equation}

\author{ Xi-zhong Liu, Jun Yu, Bo Ren }

\affiliation{Institute of Nonlinear Science, Shaoxing University, Shaoxing 312000, China}
\begin{abstract}
The residual symmetry coming from truncated Painlev\'{e} expansion of KP equation is nonlocal, which is localized in this paper by introducing multiple new dependent variables. By using the standard Lie group approach, the symmetry reduction solutions for KP equation is obtained based on the general form of Lie point symmetry for the prolonged system. In this way, the interaction solutions between solitons and background waves is obtained, which is hard to study by other traditional methods.
\end{abstract}



\maketitle
\section{Introduction}
In nonlinear science, there exist many effective ways to derive exact solutions including the soliton solutions and different wave solutions for integrable systems. Among which, the symmetry analysis plays an important role. The application of symmetry group in dealing with differential equations can date back to Sophus Lie \cite{lie} and various symmetry methods have been developed to find exact solutions for different PDEs since then. As we know, once the Lie point symmetry group for a nonlinear system is obtained, one can reduce the system by using both the classical and nonclassical Lie group methods \cite{olverbluman} and the symmetry reduction solutions is consequently derived out. The Darboux transformations (DT) and the B\"{a}cklund transformations (BT) are two other effective methods \cite{guchh}, from which, in principle, one can obtain new solutions from the known ones. However, one can only get multiple soliton solutions by taking the seed solution as constant. For the seed solutions taken as nonconstant nonlinear waves, it is difficulty to find new solutions for nonlinear systems by using BT or DT.

Painlev\'{e} analysis is an important way to investigate the integrability of nonlinear systems. For any Painlev\'{e} integrable system, there exist a truncated expansion which relate one solution to another, so it is a particular form of BT. In many cases, the BT for integrable system have been obtained by various ways, while the corresponding symmetry is rarely studied. To study the BT related symmetry and use it construct exact solutions of nonlinear system by symmetry reduction method would open an new window for study the properties of integrable systems in terms of symmetry theory. However, the BT related symmetries are nonlocal \cite{97lou}, which is hard to used directly in constructing exact solutions of nonlinear systems. To concur that obstacle, one direct way is to localize it by introducing new dependent variables so that the BT related symmetry become localized in the new system \cite{louhu}. Recently, it is found that symmetry related to the Painlev\'{e} truncated expansion is just the residue with respect to the singular manifold, which is called residual symmetry \cite{xiaonan}. Here, in this paper, we just consider the residual symmetry of KP (Kadomtsev-Petviashvili) equation and use it to construct interacting exact solutions between solitons and different nonlinear waves, which is hard to obtain by other approaches.

The KP equation 
\begin{equation}\label{kp}
(u_t-6uu_x+u_{xxx})_x+3u_{yy}=0
\end{equation}
was firstly derived to study the evolution of long ion-acoustic waves of small amplitude propagating in plasmas under the effect of long transverse perturbations \cite{kp}. The KP equation was widely considered as a natural extension of the classical KdV equation to two spatial dimensions, and has wide applications in almost all physical fields such as nonlinear optics \cite{pel}, ion-acoustic waves in plasmas, ferromagnetics, Bose-Einstein condensation and string theory.

In order to extend applicability of nonlocal symmetry to obtain explicit solutions of
KdV equation, we introduce another two auxiliary variables v and g to form a prolonged
system with u and u1, so that the original nonlocal symmetry can be transformed to a Lie
point symmetry of the new equations system.

and two kinds of novel similarity reductions.

The paper is organized as follows. In section 2, we first obtain the residual symmetry of KP equation by using the Painlev\'{e} truncated expansion method then localize it by introducing multiple new dependent variables. For the enlarged KP system, the finite group transformation theorem is consequently obtained by using Lie's first principle. In section 3, the general form of Lie point symmetry group including the localized residual symmetry for the enlarged KP system is obtained. The similarity reductions for the system is also considered according to the standard Lie point symmetry approach with some special explicit solutions are given. The last section is devoted to a short summary and discussion.

\section{new B\"{a}cklund transformation related to residual symmetry }

It is well known that the KP equation is Painlev\'{e} integrable, which means there exist a truncated expansion
\begin{equation}\label{genpain}
u=\sum_{i=0}^{\alpha}{u_i\phi^{i-\alpha}},
\end{equation}
with $\phi$ being the singular manifold and $\alpha$ being a positive integer.
Substituting \eqref{genpain} into Eq. \eqref{kp} and balancing the nonlinear term and dispersion term, we have the truncated Painlev\'{e} expansion in the form
\begin{equation}\label{pain1}
u=\frac{u_0}{\phi^2}+\frac{u_1}{\phi}+u_2.
\end{equation}
Now, substituting Eq. \eqref{pain1} into KP equation \eqref{kp}, we have
\begin{multline}\label{subspain1}
3u_{2yy}-6u_{2x}^2-6u_2u_{2xx}+u_{2xxxx}+u_{2xt}+\phi^{-1}(3u_{1yy}-6u_1u_{2xx}
+u_{1xt}-6u_2u_{1xx}\\-12u_{1x}u_{2x}+u_{1xxxx})+\phi^{-2}(-4u_{1x}\phi_{xxx}
-u_1\phi_{xt}-u_1\phi_{xxxx}-6u_{1xx}\phi_{xx}-6u_{1y}\phi_y+6u_2u_1\phi_{xx}\\-u_{1x}\phi_t
+12u_1\phi_xu_{2x}-3u_1\phi_{yy}+3u_{0yy}+12u_2u_{1x}\phi_x-6u_0u_{2xx}+u_{0xt}-6u_2u_{0xx}
-6u_1u_{1xx}-4u_{1xxx}\phi_x\\-12u_{0x}u_{2x}+u_{0xxxx}-u_{1t}\phi_x-6u_{1x}^2)
+\phi^{-3}(-12u_{0xx}\phi_{xx}-12u_2u_1\phi_x^2+6u_1\phi_y^2+8u_1\phi_x\phi_{xxx}
+12u_{1xx}\phi_x^2\\+24u_2u_{0x}\phi_x+24u_{1x}\phi_x\phi_{xx}-8u_{0xxx}\phi_x+24u_1u_{1x}
\phi_x-2u_{0t}\phi_x-2u_{0x}\phi_t+6u_1^2\phi_{xx}-6u_0\phi_{yy}-6u_0u_{1xx}\\+24u_0\phi_x
u_{2x}-8u_{0x}\phi_{xxx}+6u_1\phi_{xx}^2-2u_0\phi_{xxxx}-12u_{0x}u_{1x}+12u_2u_0\phi_{xx}
+2u_1\phi_t\phi_x-2u_0\phi_{xt}-12u_{0y}\phi_y\\-6u_1u_{0xx})+\phi^{-4}(-6u_{0x}^2+72u_{0x}
\phi_x\phi_{xx}+24u_0\phi_x\phi_{xxx}+36u_{0xx}\phi_{x}^2-36u_1\phi_x^2\phi_{xx}
-18u_1^2\phi_x^2+36u_0u_{1x}\phi_x\\+18u_0\phi_{xx}^2-36u_2u_0\phi_x^2+6u_0\phi_t\phi_x
+36u_1u_{0x}\phi_x-24u_{1x}\phi_x^3+18u_0u_1\phi_{xx}+18u_0\phi_y^2-6u_0u_{0xx})
\\+\phi^{-5}(12u_0^2\phi_{xx}-72u_0u_1\phi_x^2-96u_{0x}\phi_x^3-144u_0\phi_x^2\phi_{xx}
+48u_0u_{0x}\phi_x+24u_1\phi_x^4)+\phi^{-6}\big[6u_0\phi_x^2(2\phi_x^2-u_0)\big].
\end{multline}
For the arbitrariness of $\phi$, the coefficients of different powers of $\phi$ should equal to zero. By vanishing the coefficients of $\phi^0$ and $\phi^{-1}$ in Eq. \eqref{subspain1}, it is interesting to find that $u_1$ satisfies the linearized  KP equation with $u_2$ as a solution. So, $u_1$ is a symmetry of KP equation, which is called residual symmetry.

Vanishing the coefficients of $\phi^{-6}$, $\phi^{-5}$ and $\phi^{-4}$ in \eqref{subspain1}, we have
\begin{equation}\label{u0v0}
u_0 = 2\phi_x^2,\quad u_1 = -2\phi_{xx}
\end{equation}
and
\begin{equation}\label{u2}
u_2 =\frac{1}{6}\phi_x^{-2}(-3\phi_{xx}^2+4\phi_x\phi_{xxx}+\phi_x\phi_t+3\phi_y^2).
\end{equation}
Vanishing the coefficient of $\phi^{-3}$ or $\phi^{-2}$ in Eq. \eqref{subspain1} and using Eqs. \eqref{u0v0}, \eqref{u2}, we get the Schwarz form of KP equation
\begin{equation}\label{schwarzform}
\bigg[\frac{\phi_t}{\phi_x}+\{\phi;x\}+\frac{3}{2}\frac{\phi_y^2}{\phi_x^2}\bigg]_x
+3\bigg(\frac{\phi_y}{\phi_x}\bigg)_y=0,
\end{equation}
with $\{\phi;x\}=\frac{\phi_{xxx}}{\phi_x}-\frac{3}{2}\frac{\phi_{xx}^2}{\phi_x^2}$.
The Schwarz form \eqref{schwarzform} is invariant under the M\"{o}bious transformation
\begin{equation}
\phi\to \frac{a_1\phi+b_1}{a_2\phi+b_2},\, a_1a_2 \neq b_1b_2,
\end{equation}
which means Eq. \eqref{schwarzform} possess three symmetries $\sigma_{\phi}=d_1$,\ $\sigma_{\phi}=d_2\phi$ and
\begin{equation}\label{symmob}
\sigma_{\phi}=d_3\phi^2
\end{equation}
with arbitrary constants $d_1,\,d_2$ and $d_3.$ It is obviously that the residual symmetry $\sigma^u=u_1$ is linked with the M\"{o}bious transformation symmetry \eqref{symmob} by the linearized equation of \eqref{u2}.

Clearly, the residual symmetry
\begin{equation}\label{residualsym}
\sigma_u=-2\phi_{xx}
\end{equation}
with $\phi$ satisfying Eq. \eqref{schwarzform} is nonlocal, which is just the generator of the BT \eqref{pain1}. 

The residual symmetry of KP equation is localized as a Lie point symmetry 
\begin{subequations}\label{pointsy1}
\begin{equation}
\sigma_{u} =h,
\end{equation}
\begin{equation}
\sigma_{g} = \phi g,
\end{equation}
\begin{equation}
\sigma_h = g^2+\phi h,
\end{equation}
\begin{equation}
\sigma_{\phi} = \frac{\phi^2}{2},
\end{equation}
\end{subequations}
for the pronged system
\begin{subequations}\label{pro}
\begin{equation}\label{kp}
(u_t-6uu_x+u_{xxx})_x+3u_{yy}=0
\end{equation}
\begin{equation}\label{u2u}
u =\frac{1}{6}\phi_x^{-2}(-3\phi_{xx}^2+4\phi_x\phi_{xxx}+\phi_x\phi_t+3\phi_y^2).
\end{equation}
\begin{equation}\label{gh1}
g=\phi_x,
\end{equation}
\begin{equation}\label{gh2}
 h= g_x,
\end{equation}
\end{subequations}
which can be easily verified by substituting Eq. \eqref{pointsy1} into the linearized equations of \eqref{pro} as
\begin{subequations}\label{linear}
\begin{equation}
\sigma_{\phi,x}-\sigma_g=0,
\end{equation}
\begin{equation}
\sigma_{g,x}-\sigma_h=0,
 \end{equation}
\begin{equation}\nonumber
\phi_x\sigma_{\phi,x}\phi_t+6\sigma_u\phi_x^3-(\sigma_{\phi,t}+4\sigma_{\phi,xxx})\phi_x^2
+2(3\phi_{xx}\sigma_{\phi,xx}+2\sigma_{\phi,x}\phi_{xxx}-3\phi_y\sigma_{\phi,y})\phi_x
\end{equation}
\begin{equation}
-6\sigma_{\phi,x}\phi_{xx}^2+6\sigma_{\phi,x}\phi_y^2=0,
\end{equation}
\begin{equation}\nonumber
(\sigma_{\phi,x}\phi_{xx}+\phi_{x}\sigma_{\phi,xx})\phi_{t}-(\sigma_{\phi,xxxx}
+\sigma_{\phi,xt}+3\sigma_{\phi,yy})\phi_{x}^2+\big[-2\phi_{xt}\sigma_{\phi,x}
\end{equation}
\begin{equation}\nonumber
+(\sigma_{\phi,t}
+4\sigma_{\phi,xxx})\phi_{xx}
+4\sigma_{\phi,xx}\phi_{xxx}-2\phi_{xxxx}\sigma_{\phi,x}-6\phi_{yy}\sigma_{\phi,x}\big]\phi_{x}
\end{equation}
\begin{equation}
-9\phi_{xx}^2\sigma_{\phi,xx}+
4\sigma_{\phi,x}\phi_{xx}\phi_{xxx}+3\sigma_{\phi,xx}\phi_{y}^2+6\phi_{xx}
\phi_{y}\sigma_{\phi,y}=0.
\end{equation}
\end{subequations}

In other words, the prolonged system \eqref{pro} has the Lie point symmetry vector
\begin{equation}\label{pointV}
V=h\partial_{u}+g\phi\partial_{g}+(g^2+h\phi)\partial_{h}
+\frac{\phi^2}{2}\partial_{\phi},
\end{equation}
which localize the residual symmetry.

Now, let us proceed to study the finite transformation form of Eq. \eqref{pointV}, which can be stated in the following theorem.

\noindent\emph{ \textbf{Theorem 1.}}
If $\{u,g,h,\phi\}$ is a solution of the prolonged system \eqref{pro}, then so is $\{\hat{u},\hat{g},\hat{h},\hat{\phi}\}$ with
\begin{subequations}\label{pointsy}
\begin{equation}
\hat{u}=\frac{(2g^2-2h\phi+u\phi^2)\epsilon^2+(4h-4u\phi)\epsilon+4u}{(\epsilon\phi-2)^2},
\end{equation}
\begin{equation}
\hat{g} = \frac{4g}{(\epsilon\phi-2)^2},
\end{equation}
\begin{equation}
\hat{h} = \frac{4h}{(\epsilon\phi-2)^2}-\frac{8\epsilon g^2}{(\epsilon\phi-2)^3},
\end{equation}
\begin{equation}
\hat{\phi} =-\frac{2\phi}{\epsilon\phi-2},
\end{equation}
\end{subequations}
with arbitrary group parameter $\epsilon$.

\emph{Proof.} Using Lie's first theorem on vector \eqref{pointV} with the corresponding
 initial condition, i.e.,
\begin{eqnarray}
\\ \frac{d\hat{u}(\epsilon)}{d\epsilon}&=& \hat{h}(\epsilon),\,\quad \hat{u}(0)=u,\\
\frac{d\hat{g}(\epsilon)}{d\epsilon}&=& \hat{\phi}(\epsilon)\hat{g}(\epsilon),\,\quad \hat{g}(0)=g,\\
\frac{d\hat{h}(\epsilon)}{d\epsilon}&=&\hat{g}(\epsilon)^2+\hat{\phi}(\epsilon)\hat{h}(\epsilon),\,\quad \hat{h}(0)=h,\\
\frac{d\hat{\phi}(\epsilon)}{d\epsilon}&=&\frac{\hat{\phi}(\epsilon)^2}{2},\,\quad \hat{\phi}(0)=\phi.
\end{eqnarray}
Solving the above equations, we find the solutions \eqref{pointsy}, thus the theorem is
proved.
\section{new symmetry reductions of KP equation}
Let us now proceed to seek the Lie point symmetry of the prolonged KP system \eqref{pro} in the general form
 \begin{equation}\label{vectorv1}
V=X\frac{\partial}{\partial x}+T\frac{\partial}{\partial
t}+U\frac{\partial}{\partial u}+G\frac{\partial}{\partial g}+H\frac{\partial}{\partial h}+\Phi\frac{\partial}{\partial \phi},
\end{equation}
which means that \eqref{pro} is invariant under the following transformation
\begin{equation}
\{x,t,u,g,h,\phi\} \rightarrow \{x+\epsilon X,t+\epsilon T,u+\epsilon U,g+\epsilon G,h+\epsilon H,\phi+\epsilon \Phi\}
\end{equation}
with a infinitesimal parameter $\epsilon$.
Equivalently, the symmetry in the form \eqref{vectorv1} can be written as a function form
\begin{subequations}\label{sigmasy}
\begin{equation}
\sigma_{u} = Xu_{x}+Tu_{t}-U,
\end{equation}
\begin{equation}
\sigma_{g} = Xg_{x}+Tg_{t}-G,
\end{equation}
\begin{equation}
\sigma_h = Xh_{x}+Th_{t}-H,
\end{equation}
\begin{equation}
\sigma_{\phi} = X\phi_{x}+T\phi_{t}-\Phi.
\end{equation}
\end{subequations}
Substituting Eq. \eqref{sigmasy} into Eq. \eqref{linear} and eliminating $u_{x}, g_{x}, g_t, h_t, \phi_x, \phi_t$ and $u_{yy}$ in terms of the prolonged system \eqref{pro}, we get more than 900 determining equations for the functions $X, Y, T, U, G, H$ and $\Phi$. Calculated by computer algebra, the general solutions are
\begin{eqnarray}\label{sol}
\nonumber&&X=\frac{1}{2}f_{1t}x-\frac{1}{12}f_{1tt}y^2-\frac{1}{6}f_{2t}y+f_3, Y=f_{1t}y+f_2, T=\frac{3}{2}f_1+c_1,\\\nonumber&&U =-\frac{1}{12}f_{1tt}x+\frac{1}{72}f_{1ttt}y^2+\frac{1}{36}f_{2tt}y-\frac{1}{6}f_{3t}-uf_{1t}-hc_2,
\\\nonumber&&G = -\frac{1}{2}gf_{1t}-c_2g\phi+c_3g, H=-c_2g^2-hf_{1t}-c_2h\phi+c_3h,\\&&\Phi =-\frac{1}{2}c_2\phi^2+c_3\phi+c_4.
\end{eqnarray}
where $f_1, f_2, f_3$ are arbitrary functions of $t$ and $c_1, c_2, c_3, c_4$ are arbitrary constants.
 
It is noted that, if we take $f_1=f_2=f_3=c_1=c_3=c_4=0$ and $c_2=1$ in Eq. \eqref{sol}, the obtained general form of symmetry degenerated into the special form in Eq. \eqref{pointV}. The time related $x$ translation invariance and $t$ translation invariance symmetries can also be obtained by setting $f_1=f_2=c_1=c_2=c_3=c_4=0$ and $f_1=f_2=f_3=c_2=c_3=c_4=0, c_1=1$, respectively.

Consequently, the symmetries in \eqref{sigmasy} can be written as
\begin{eqnarray}\label{sigmauvf}
\nonumber\sigma_{u}&=&(\frac{1}{2}f_{1t}x-\frac{1}{12}f_{1tt}y^2-\frac{1}{6}f_{2t}y+f_3)u_x
+(f_{1t}y+f_2)u_y+(\frac{3}{2}f_1+c_1)u_t\\&&\nonumber+\frac{1}{12}f_{1tt}x-\frac{1}{72}f_{1ttt}y^2
-\frac{1}{36}f_{2tt}y+\frac{1}{6}f_{3t}+uf_{1t}+hc_2,\\
\nonumber\sigma_{g}&=&(\frac{1}{2}f_{1t}x-\frac{1}{12}f_{1tt}y^2-\frac{1}{6}f_{2t}y+f_3)g_x
+(f_{1t}y+f_2)g_y+(\frac{3}{2}f_1+c_1)g_t\\&&\nonumber+\frac{1}{2}gf_{1t}+c_2g\phi-c_3g,\\
\nonumber\sigma_h&=&(\frac{1}{2}f_{1t}x-\frac{1}{12}f_{1tt}y^2-\frac{1}{6}f_{2t}y+f_3)h_x
+(f_{1t}y+f_2)h_y+(\frac{3}{2}f_1+c_1)h_t\\&&\nonumber+c_2g^2+hf_{1t}+c_2h\phi-c_3h,\\
\nonumber\sigma_{\phi}&=&(\frac{1}{2}f_{1t}x-\frac{1}{12}f_{1tt}y^2-\frac{1}{6}f_{2t}y+f_3)\phi_x
+(f_{1t}y+f_2)\phi_y+(\frac{3}{2}f_1+c_1)\phi_t\\&&+\frac{1}{2}c_2\phi^2-c_3\phi-c_4.
\end{eqnarray}

Similarity reduction solutions of the enlarged system can be found by setting $\sigma_u=\sigma_g=\sigma_h=\sigma_{\phi}=0$ in Eq. \eqref{sigmauvf}, which is equivalent to solve the corresponding characteristic equations
\begin{equation}\label{chac}
\frac{dx}{X}=\frac{dy}{Y}=\frac{dt}{ T}=\frac{du}{U}=\frac{dg}{G}=\frac{dh}{H}
=\frac{d\phi}{\Phi}.
\end{equation}

In the following part of the paper, three nontrivial cases are considered regarding with the symmetry reduction.

\noindent \textbf{Case 1}. $f_1\neq0$ and $c_1\neq0$.

In this case, without loss of generality, we rewrite the arbitrary functions $f_2$ and $f_3$ as
\begin{equation}
f_2=(3f_1+2c_1)^{\frac{5}{3}}m_{1t}
\end{equation}
\begin{equation}
f_3 =-\frac{1}{3}(3f_1+2c_1)^{\frac{4}{3}}\big[(3f_1+2c_1)m_{1t}^2-m_{2t}\big]
\end{equation}

with $m_1$ and $m_2$ being arbitrary functions of $t$.

Now, by solving Eq. \eqref{chac}, we get
\begin{eqnarray}\label{redusol01}
\phi&=&\frac{\Delta\tanh\big[\Delta(m_3+\Phi)\big]+c_3}{c_2},\,\Delta=\sqrt{2c_2c_4+c_3^2},\\
\label{redusol02}g&=&\frac{G}{(3f_1+2c_1)^{1/3}\cosh\big[\Delta(m_3+\Phi)\big]^2},\\
\label{redusol03}h&=&-\frac{2c_2G^2\sinh\big[\Delta(m_3+\Phi)\big]-H\Delta\cosh\big[\Delta(m_3+\Phi)\big]}
{(3f_1+2c_1)^{2/3}\Delta\cosh\big[\Delta(m_3+\Phi)\big]^3},\\
\label{redusol04}u&=&\frac{1}{36(3f_1+2c_1)^{2/3}\Delta^2\bigg[1+\tanh^2\big[\frac{1}{2}\Delta(m_3+\Phi)\big]\bigg]^2}
\left\{\Delta^2\Omega\tanh^4\bigg[\frac{1}{2}\Delta(m_3+\Phi)\bigg]\right.\nonumber\\&&-144c_2H\Delta
\tanh^3\bigg[\frac{1}{2}\Delta(m_3+\Phi)\bigg]+\bigg[2\Delta^2(\Omega+36U)+288c_2^2G^2\bigg]\nonumber\\&&\left.\tanh^2\bigg[\frac{1}{2}\Delta(m_3+\Phi)\bigg]
-144c_2H\Delta\tanh\bigg[\frac{1}{2}\Delta(m_3+\Phi)\bigg]+\Delta^2(\Omega+36U)\right\},
\end{eqnarray}
where $\Phi, U, G$ and $H$ are group invariants functions of the similarity variables $\xi$ and $\eta$, which can be expressed as
\begin{equation}\label{rxi}
\xi=\frac{f_{1t}}{6(3f_1+2c_1)^{\frac{4}{3}}}y^2+\frac{1}{3}(3f_1+2c_1)^{\frac{1}{3}}m_{1t}y
+\frac{x}{(3f_1+2c_1)^{\frac{1}{3}}}-\frac{2}{3}m_2,
\end{equation}
\begin{equation}\label{reta}
\eta=\frac{y}{(3f_1+2c_1)^{\frac{2}{3}}}-2m_1.
\end{equation}
In Eq. \eqref{redusol04}, we had wrote
\begin{equation}
m_3=\int{\frac{1}{3f_1+2c_1}dt}
\end{equation}
\begin{multline}
\Omega = -\frac{6f_{1t}}{(3f_1+2c_1)^{\frac{1}{3}}}x+\bigg[\frac{f_{1tt}}{(3f_1+2c_1)^{\frac{1}{3}}}-\frac{2f_{1t}^2}
{(3f_1+2c_1)^{\frac{4}{3}}}\bigg]y^2+2\bigg[3(3f_1+2c_1)^{\frac{1}{3}}m_{1t}f_{1t}\\+(3f_1+2c_1)^{\frac{4}{3}}
m_{1tt}\bigg]y+2(3f_1+2c_1)^2m_{1t}^2-4(3f_1+2c_1)m_{2t}
\end{multline}
for simplicity.

Substituting Eqs. \eqref{redusol01}, \eqref{redusol02}, \eqref{redusol03} and \eqref{redusol04} into the prolonged system, we get the corresponding symmetry reduction equations
\begin{equation}\label{redgphi}
\Delta^2\Phi_{\xi}-c_2G=0,
\end{equation}
\begin{equation}\label{redhphi}
\Delta^2\Phi_{\xi\xi}-c_2H=0,
\end{equation}
\begin{equation}\label{reduphi}
-3\Phi_{\eta}^2+8\Delta^2\Phi_{\xi}^4+6\Phi_{\xi}^2U+(-1-4\Phi_{\xi\xi\xi})\Phi_{\xi}
+3\Phi_{\xi\xi}^2=0
\end{equation}
and
\begin{multline}\label{redphi}
-3\Phi_{\xi\xi}\Phi_{\eta}^2-4\Delta^2\Phi_{\xi\xi}\Phi_{\xi}^4+(3\Phi_{\eta\eta}
+\Phi_{\xi\xi\xi\xi})\Phi_{\xi}^2-(1+4\Phi_{\xi\xi\xi})\Phi_{\xi\xi}\Phi_{\xi}+3
\Phi_{\xi\xi}^3=0
\end{multline}

By solving out $G, H$ and $U$ from Eqs. \eqref{redgphi}, \eqref{redhphi}, \eqref{reduphi} and then substituting them into Eq. \eqref{redusol04}, we have the following theorem

\noindent\textbf{\emph{Theorem 2}} If $\Phi$ is a solution of the reduction equation \eqref{redphi},
then $u$ given by
\begin{multline}\label{solexu}
u=\frac{1}{36(3f_1+2c_1)^{2/3}\bigg[1+\tanh^2\big[\frac{1}{2}\Delta(m_3+\Phi)\big]\bigg]^2\Phi_{\xi}^2}
\bigg[(18\Phi_{\eta}^2-48\Delta^2\Phi_{\xi}^4+\Omega\Phi_{\xi}^2\\+(24\Phi_{\xi\xi\xi}
+6)\Phi_{\xi}-18\Phi_{\xi\xi}^2)\tanh^4\big[\frac{1}{2}\Delta(m_3+\Phi)\big]-
144\Phi_{\xi}^2\Phi_{\xi\xi}\Delta\tanh^3\big[\frac{1}{2}\Delta(m_3+\Phi)\big]\\+2(18\Phi_{\eta}^2
+96\Delta^2\Phi_{\xi}^4+\Omega\Phi_{\xi}^2+(24\Phi_{\xi\xi\xi}+6)\Phi_{\xi}-18
\Phi_{\xi\xi}^2)\tanh^2\big[\frac{1}{2}\Delta(m_3+\Phi)\big]\\-144\Phi_{\xi}^2\Phi_{\xi\xi}\Delta
\tanh\big[\frac{1}{2}\Delta(m_3+\Phi)\big]+18\Phi_{\eta}^2-48\Delta^2\Phi_{\xi}^4+\Omega\Phi_{\xi}^2\\+(24\Phi_{\xi\xi\xi}
+6)\Phi_{\xi}-18\Phi_{\xi\xi}^2\bigg]
\end{multline}
is a solution of KP equation.

From Eq. \eqref{solexu}, we can obtain some interesting solutions even $\Phi$ is a trivial solution of Eq. \eqref{redphi}. As an illustration, we take the solution of \eqref{redphi} as $\Phi=\xi+\eta$ and then we find
\begin{multline}
u=\frac{1}{36(3f_1+2c_1)^{2/3}\bigg[1+\tanh^2\big[\frac{1}{2}\Delta(m_3+\xi+\eta)\big]\bigg]^2}
\bigg[(24-48\Delta^2+\Omega) \tanh^4\big[\frac{1}{2}\Delta(m_3+\xi+\eta)\big]\\+2(24
+96\Delta^2+\Omega)\tanh^2\big[\frac{1}{2}\Delta(m_3+\xi+\eta)\big]+24-48\Delta^2+\Omega\bigg]
\end{multline}
with $\xi, \eta$ expressed by Eqs. \eqref{rxi} and \eqref{reta} is a nontrivial solution of KP equation.

It is obvious that Eq. \eqref{solexu} can be considered as an interaction solution of soliton with the background wave.

\noindent\textbf{Case 2.} $f_1=c_1=0$ and $f_2\neq0$.

Similarly to case 1, we get the second type symmetry reduction solutions for the enlarged KP system
\begin{eqnarray}\label{redusol001}
\phi&=&\frac{1}{c_2}\left\{c_3-\Delta\tanh\bigg[\frac{\Delta(6f_3-f_{2t}y-3f_{2t}f_2\phi')}{2f_{2t}f_2}\bigg]\right\},\\
\label{redusol002}g&=&\frac{g'}{\cosh^2\bigg[\frac{\Delta(6f_3-f_{2t}y-3f_{2t}f_2\phi')}{2f_{2t}f_2}\bigg]},\\
\label{redusol003}h&=&\frac{1}{\Delta\cosh^3\bigg[\frac{\Delta(6f_3-f_{2t}y-3f_{2t}f_2\phi')}{2f_{2t}f_2}\bigg]}
\left\{2c_2g'^2\sinh\bigg[\frac{\Delta(6f_3-f_{2t}y-3f_{2t}f_2\phi')}{2f_{2t}f_2}\bigg]\right.\\&&\left.+\Delta h'\cosh\bigg[\frac{\Delta(6f_3-f_{2t}y-3f_{2t}f_2\phi')}{2f_{2t}f_2}\bigg]\right\},\\
\label{redusol004}u&=&\frac{2c_2^2}{\Delta^2}g'^2\tanh^2\bigg[\frac{\Delta(6f_3-f_{2t}y-3f_{2t}f_2\phi')}{2f_{2t}f_2}\bigg]
+\frac{2c_2}{\Delta}h'\tanh\bigg[\frac{\Delta(6f_3-f_{2t}y-3f_{2t}f_2\phi')}{2f_{2t}f_2}\bigg]
\nonumber\\&&-\frac{1}{72f_2f_{2t}^2}\left\{\big[(12y+36\phi' f_2)f_{3t}+(3\phi' f_2-y)(y+3\phi' f_2)f_{2tt}-72u'f_2\big]f_{2t}^2\right.\nonumber\\&&\left.-36(2f_{3t}+f_{2tt}f_2\phi')f_{2t}f_3+36f_{2tt}f_3^2\right\},
\end{eqnarray}
where $u'\equiv u'(x',t), g'\equiv g'(x',t), h'\equiv h'(x',t), \phi'\equiv\phi'(x',t)$ with
\begin{equation}
x'=-x-\frac{f_{2t}}{12f_2}y^2+\frac{f_3}{f_2}y.
\end{equation}

The corresponding symmetry reduction equations can be obtained by substituting Eqs. \eqref{redusol001}, \eqref{redusol002}, \eqref{redusol003}, \eqref{redusol004}, into prolonged KP system \eqref{pro}, the result is
\begin{equation}\label{2gphi}
3\Delta^2\phi_{x'}'+2c_2g'=0,
\end{equation}
\begin{equation}\label{2hphi}
3\Delta^2\phi_{x'x'}'-2c_2h'=0,
\end{equation}
\begin{multline}\label{2uphi}
12\phi_{x'}'f_{2t}^2\phi_t'f_2^2+216f_2^2\Delta^2f_{2t}^2\phi_{x'}'^4-9\big[(4f_2^2f_{3t}
\phi'+f_2^3f_{2tt}\phi'^2-8f_2^2u'+4f_3^2)f_{2t}^2-\\4(2f_2f_{3t}f_3+f_2^2f_{2tt}
\phi' f_3)f_{2t}+4f_2f_{2tt}f_3^2\big]\phi_{x'}'^2-24(f_{2t}f_2f_{3t}-f_{2tt}f_2f_3+2f_{2t}^2f_2^2
\phi_{x'x'x'}')\phi_{x'}'\\-4(1-9f_2^2\phi_{x'x'}'^2)f_{2t}^2=0
\end{multline}
and
\begin{multline}\label{2redphi}
6\phi_{x'}'\phi_{x'x'}'f_{2t}^2\phi_t'f_2^2-54f_2^2\Delta^2\phi_{x'x'}'
f_{2t}^2\phi_{x'}'^4-3\phi_{x'}'^3f_{2t}^3f_2-6(\phi_{x't}'f_2^2-f_2^2\phi_{x'x'x'x'}'
)f_{2t}^2\phi_{x'}'^2\\-12(\phi_{x'x'}'f_{2t}f_2f_{3t}-\phi_{x'x'}'f_{2tt}f_2f_3+2
f_{2t}^2f_2^2\phi_{x'x'}'\phi_{x'x'x'}')\phi_{x'}'+2(9f_2^2\phi_{x'x'}'^3-
\phi_{x'x'}')f_{2t}^2=0
\end{multline}

Similar to Theorem 2, solving out $g', h'$ and $\phi'$ from Eqs. \eqref{2gphi}, \eqref{2hphi}, \eqref{2uphi} and substituting them into Eq. \eqref{redusol004}, we have the following theorem

\noindent\textbf{\emph{Theorem 3}}. If $\phi'$ is a solution of the symmetry reduction equation
\eqref{2redphi}, then $u$ given by
\begin{multline}
u=-\frac{9}{2}\Delta^2\phi_{x'}'^2\tanh\bigg[\frac{\Delta(6f_3-f_{2t}y-3f_{2t}f_2\phi')}
{2f_{2t}f_2}\bigg]+3\Delta\phi_{x'x'}'\tanh\bigg[\frac{\Delta(6f_3-f_{2t}y-3f_{2t}f_2\phi')}
{2f_{2t}f_2}\bigg]\\-\frac{1}{72f_2^2f_{2t}^2\phi_{x'}'^2}\bigg[12\phi_{x'}'f_{2t}^2
\phi_{t}'f_2^2+216f_2^2\Delta^2f_{2t}^2\phi_{x'}'^4+(
12f_2f_{3t}y-36f_3^2-f_2f_{2tt}y^2)f_{2t}^2\phi_{x'}'^2
\\-24(f_{2t}f_2f_{3t}-f_{2tt}f_2f_3+2f_{2t}^2f_2^2\phi_{x'x'x'}')\phi_{x'}'-4(1-9f_2^2\phi_{x'x'}'^2)f_{2t}^2\bigg]
\end{multline}
is a solution of the KP equation.

\noindent\textbf{Case 3} $f_1=f_2=c_1=0$.

In this case, executing the similar procedure as in case 1 and 2, we can find the third type of group invariant solutions
\begin{eqnarray}\label{redusol}
g&=&-\frac{g''}{\cosh^2\big[\frac{\Delta(x+\phi'')}{2f_3}\big]},\\
h&=&\frac{1}{\Delta\cosh^3\big[\frac{\Delta(x+\phi'')}{2f_3}\big]\bigg[\sinh\big[\frac{\Delta(x+\phi'')}{2f_3}\big]
+\cosh\big[\frac{\Delta(x+\phi'')}{2f_3}\big]\bigg]}\left\{2c_2g''^2\right.\\&&\left.-h''\Delta\cosh^2\big[\frac{\Delta(x+\phi'')}{2f_3}\big]
-
h''\Delta\sinh\big[\frac{\Delta(x+\phi'')}{2f_3}\big]\cosh\big[\frac{\Delta(x+\phi'')}{2f_3}\big]\right\},\\
\phi&=&\frac{1}{c_2}\bigg[c_3+\tanh\big[\frac{\Delta(x+\phi'')}{2f_3}\big]\Delta\bigg],\\
u&=&\frac{2c_2^2}{\Delta^2}g''^2\tanh^2\big[\frac{\Delta(x+\phi'')}{2f_3}\big]-\frac{2c_2}{\Delta^2}
(2c_2g''^2-\Delta h'')\tanh\big[\frac{\Delta(x+\phi'')}{2f_3}\big]\nonumber\\&&-\frac{f_{3t}}{6f_3}x+u''\label{003u}
\end{eqnarray}
with group invariant functions $u''\equiv u''(y,t), g''\equiv g''(y,t), h''\equiv h''(y,t), \phi''\equiv \phi''(y,t)$.

The corresponding symmetry reduction equations are
\begin{equation}\label{3redg}
\Delta^2+2c_2f_3g''=0
\end{equation}
\begin{equation}\label{3redh}
-\Delta^3+2c_2f_3^2h''=0
\end{equation}
\begin{equation}\label{3reduphi}
-2\Delta^2+3f_3^2\phi_{y}''^2-6u''f_3^2-f_3f_{3t}\phi''+f_3^2\phi_t=0
\end{equation}
\begin{equation}\label{3redphi}
-3f_3\phi''_{yy}+f_{3t}=0
\end{equation}

Again, solving out $g'', h''$ and $u''$ by Eqs. \eqref{3redg}, \eqref{3redh}, \eqref{3reduphi} and substituting them into Eq. \eqref{003u}, then we have the following theorem

\noindent\textbf{\emph{Theroem 4}}. If $\phi''$ is a solution of the symmetry reduction equation \eqref{3redphi}, then $u$ given by
\begin{equation}
u =\frac{\Delta^2}{2f_3^2}\tanh^2\big[\frac{\Delta(x+\phi'')}{2f_3}\big]
+\frac{1}{2}\phi_{y}''^2+\frac{1}{6}\phi_t''-\frac{f_{3t}}{6f_3}(x+\phi'')
-\frac{\Delta^2}{3f_3^2}
\end{equation}
is a solution of KP equation.

\section{Conclusion and discussion}
In summary, the residual symmetry coming from the Painlev\'{e} truncated expansion of KP equation is localized by introducing multiple new dependent variables. For the prolonged KP system, the new B\"{a}cklund transformation is derived by using Lie's first principle. The general form of Lie point symmetry for the prolonged system is found, from which, it can be found that the residual symmetry is included as a special ones. The Lie point symmetry group for the prolonged KP system covers the corresponding one for the original KP equation, so many more rich interaction solutions can be found by using the classical Lie group approach. From the form of symmetry reduction solution for the KP equation, we see that the residual symmetry is to add an additional soliton to the background wave.
In other words, the interaction solution between soliton and back ground wave is successfully obtained. We also give some special explicit solutions under certain constraints.

In this paper, we have shown that the residual symmetry can be used to construct interaction solutions for nonlinear equations. It is meaningful to try use other form of nonlocal symmetries, which can be found from B\"{a}cklund transformation, the bilinear forms and negative hierarchies, the nonlinearizations \cite{cao,cheng} and self-consistent sources \cite{zeng} etc. to construct new exact solutions. Despite of the bright prospect, there still exist some open questions deserved to probe. First, it still remains unclear what kind of nonlocal symmetries must have close prolongations and can be applied to construct exact solutions. Second, different nonlocal symmetries leads to different symmetry reduction solutions, for example the residual symmetry and the DT related symmetry, so what is the link between them? can the nonlocal symmetries be classified so that symmetries in one kind leads to essentially the same kind of interaction solution? All in all, applying nonlocal symmetries to construct exact solutions for nonlinear systems is adventurous but meaningful.

\begin{acknowledgments}
The authors are in debt to thank Prof. S.Y. Lou for his valuable comments and suggestions. This work was supported by the National Natural Science Foundation of China under Grant Nos. 10875078 and 11305106, the Natural Science Foundation of Zhejiang Province of China Grant Nos. Y7080455 and LQ13A050001.
\end{acknowledgments}

\end{document}